\documentclass[doublecol]{epl2}
\title{Gravity Localization and Effective Newtonian Potential for Bent Thick Branes}
\shorttitle{Gravity Localization and Effective Newtonian Potential for Bent Thick Branes} %Insert here a short version of the title if it exceeds 70 characters

\author{Heng Guo\thanks{E-mail: \email{guoh2009@lzu.edu.cn}}
 \and Yu-Xiao Liu\thanks{Corresponding author. E-mail: \email{liuyx@lzu.edu.cn}}
 \and Shao-Wen Wei\thanks{E-mail: \email{weishw08@lzu.cn}}
 \and Chun-E Fu\thanks{E-mail: \email{fuche08@lzu.cn}} }
\shortauthor{H. Guo \etal}

\institute{
   Institute of Theoretical Physics,
    Lanzhou University,
    Lanzhou 730000, P. R. China
}
\pacs{04.50.-h}{Higher-dimensional gravity and other theories of gravity}
\pacs{11.27.+d}{Extended classical solutions; cosmic strings, domain walls, texture}

\abstract{
In this letter, we first investigate the gravity localization and
mass spectrum of gravity KK modes on de Sitter and Anti-de Sitter
thick branes. Then, the effective Newtonian gravitational potentials
for these bent branes are discussed by the two typical examples. The
corrections of the Newtonian potential turns out to be $\Delta
U(r)\sim 1/r^{2}$ at small $r$ for both cases. These corrections are
very different from that of the Randall-Sundrum brane model $\Delta
U(r)\sim 1/r^{3}$.}

\begin{document}

\maketitle

\section{Introduction}

The brane world models can supply new insights for solving gauge
hierarchy and the cosmological constant problem
\cite{RubakovPLB1983136,rs,Lykken,ADD,CosmConst}. The key idea of
brane world is that gravity is free to propagate in all dimensions,
while all matter fields are confined to a 3-brane with no
contradiction with present gravitational experiments. In 1999, the
brane world models with warped extra dimension were proposed by
Randall and Sundrum (RS) \cite{rs}. In the famous RS brane-world
model \cite{rs}, the thickness of the brane is neglected.
Subsequently, DeWolfe, Freedman, Gubser and Karch
\cite{De_Wolfe_PRD_2000} proposed a more natural brane world model,
the so called thick-brane scenario, in which the brane is generated
by a background scalar field coupled gravity. Since then, more and
more authors have investigated the thick-brane scenario in higher
dimensional space-time \cite{Gremm_2000}-\cite{Liu_2010}. In these
models, the scalar fields do not play the role of bulk fields but
provide the ``material" of which the thick branes are made. For some
comprehensive reviews about thick branes please see Ref.
\cite{reviews}.

In brane world scenarios, there is an interesting and important
issue, i.e., the localization of gravity. Generally, the gravity
zero mode should be localized on the brane for the purpose that the
effective four-dimensional gravity should be recovered. Moreover, from
the phenomenological point of view, it is very important to seek
configurations which allow for the existence of a mass gap in the
spectrum of gravitational Kaluza-Klein (KK) excitations. The mass
gap can decide the energy scale beyond which massive modes can be
excited. From Refs. \cite{Wang_PRD,PRD0709.3552} we know that the
existence of a mass gap would in general make the localization
of the massless graviton sure. Especially, the mass gap can result in
different corrections of the Newton potential \cite{liu_0911.0269}.

In this letter, we will investigate the localization and mass
spectrum of gravity as well as the effective Newtonian potentials on de Sitter
(dS) and anti-de Sitter (AdS) branes by some typical examples. For the case of dS
branes, the potential of the corresponding Schr\"{o}dinger equation
for the gravity KK modes is a modified P\"{o}schl-Teller potential,
and we find that the correction term for the Newtonian potential at
small $r$ is $\sim 1/r^{2}$, which is different from the correction
caused by a volcano-like localized potential. For the case of AdS
branes, the localized potential of the KK modes is an infinite deep
potential, and all the KK modes are bound states. It is shown that
the correction term of the gravitational potential at small distance
for this case is also proportional to $ 1/r^{2}$.

The organization of this letter is as follows. In next section, we
review the localization of gravity. Then, we investigate the
localization and mass spectrum of gravity on dS and AdS thick branes
in the third and fourth section, respectively.
The effective gravitational potentials on bent branes are studied in the fifth
section. Finally, the conclusion is given in the last section.

\section{The model}
\label{SecModel}

We start with the following 5D action of bent thick
branes, which are generated by a real scalar field $\phi$ with a
scalar potential $V(\phi)$,
\begin{equation}
 S \!\! = \!\! \int d^5 x \sqrt{-g}\left [ \frac{1}{2\kappa_5^2}
     R-\frac{1}{2}
     \partial^M \phi \partial_N \phi - V(\phi) \right ],
\label{action}
\end{equation}
where $R$ is the 5D scalar curvature.
The line-element is assumed as
\begin{eqnarray}\label{line_z}
 ds^{2}= g_{MN}dx^{M}dx^{N}=\textrm{e}^{2A(z)}[\hat{g}_{\mu\nu}(x)dx^\mu dx^\nu + dz^2],
\end{eqnarray}
where $\textrm{e}^{2A(z)}$ is the warp factor and $z$ stands for the
extra coordinate. We suppose that $\hat{g}_{\mu\nu}$ is some general
4D metric, and $\phi=\phi(z)$.

The field equations corresponding to the action (\ref{action}) are read as
\begin{eqnarray}
{\kappa_{5}^{2}} ~\phi'^{2}~&=&3A'^{2}-3A''-\Lambda_{4},\label{dS_EinsteinEq_a} \\
{2\kappa_{5}^{2}}~V(\phi)  &=&{3}
             \textrm{e}^{-2A}(\Lambda_{4}-3A'^{2}-A''), \label{dS_EinsteinEq_b}\\
\frac{d V(\phi)}{d\phi}
         &=& \textrm{e}^{-2A}(3A'\phi'+\phi'').\label{dS_EinsteinEq_c}
\end{eqnarray}
Here we set the 4D metric satisfies $\hat{R}_{\mu\nu}=\Lambda_{4}\hat{g}_{\mu\nu}$
with $\Lambda_{4}$ the 4D cosmology constant, i.e., the 4D space-time is maximally symmetric 4D space-time. In the case of AdS$_{4}$ space, $\Lambda_{4}$ is negative, while for dS$_{4}$ space the converse holds true.
In the following sections, we will consider some explicit solutions of these equations.

Let us further consider the metric fluctuations $\delta
g_{MN}=H_{MN}=\textrm{e}^{2A(z)}h_{MN}$ of (\ref{line_z}). Under the
axial gauge $H_{5M}=0$, the total metric can be written as
\begin{eqnarray}
 ds^{2}= \textrm{e}^{2A(z)}\left[\big(\hat{g}_{\mu\nu}(x)
    +h_{\mu\nu}(x,z)\big)dx^\mu dx^\nu + dz^2\right].
\end{eqnarray}
After imposing the transverse-traceless (TT) gauge condition $h^{\mu}_{\mu}=\hat{\nabla}^{\mu}h_{\mu\nu}=0$, where $\hat{\nabla}_{\alpha}$ denotes the covariant derivative with
respect to $\hat{g}_{\mu\nu}$, the equation for the perturbation $h_{\mu\nu}$ takes following form~\cite{RandallJHEP2001}:
\begin{eqnarray}\label{dynamics_eq}
\left(\partial_{z}^{2} + 3(\partial_{z}A)\partial_{z}+
 %\Box^{(4)}
 \hat{g}^{\alpha\beta}\hat{\nabla}_{\alpha}\hat{\nabla}_{\beta}
   -\frac{2}{3}\Lambda_{4}\right)h_{\mu\nu}(x,z)=0.
\end{eqnarray}
By making use of the KK decomposition $h_{\mu\nu}(x,z) =
\textrm{e}^{-3A/2}\epsilon_{\mu\nu}(x)\Psi(z)$ with $\epsilon_{\mu\nu}(x)$ satisfying the TT condition, from Eq. (\ref{dynamics_eq}) we can get the following  4D equation: $(\hat{g}^{\alpha\beta}\hat{\nabla}_{\alpha}\hat{\nabla}_{\beta}
   -\frac{2}{3}\Lambda_{4})\epsilon_{\mu\nu}(x)=m^{2}\epsilon_{\mu\nu}(x)$, and the Schr\"{o}dinger-like equation for the 5th dimensional sector
\begin{eqnarray}\label{Schrodinger_eq}
\left[-\partial_{z}^{2}+V_{QM}(z)\right]\Psi(z)=m^{2}\Psi(z),
\end{eqnarray}
where $m$ is the mass of the KK modes. The localizing potential is read as
\begin{eqnarray}\label{VQM}
V_{QM}=\frac{3}{2}\left[\partial_{z}^{2}A+\frac{3}{2}(\partial_{z}A)^{2}\right].
\end{eqnarray}
From Refs. \cite{Csaki_NPB_2000,Brandhuber_JHEP}, we know that, in order to
localize the massless mode of gravity on a brane, the potential $V_{QM}$ should
have a well with a negative minimum inside the brane and satisfy
$V_{QM}(z\rightarrow\pm\infty)\geq 0$ when far from the brane.
%Moreover, it is desirable to obtain
%solutions in which the massless graviton is separated from the
%massive modes by a mass gap \cite{Brandhuber_JHEP} for a well
%defined localizing potential. This phenomenological aspect ensures
%the lack of arbitrary light KK excitations in the mass spectrum.
%According to quantum mechanics, if the potential has the asymptotic
%behavior $V_{QM}(z\rightarrow\pm\infty)>0$, the existence of a mass
%gap is guaranteed.

The zero mode is
$\Psi_{0}(z)\propto \textrm{e}^{\frac{3}{2}A(z)}$ \cite{RandallJHEP2001}.
%which accords with the 4D massless graviton solution
%$h_{\mu\nu}^{TT}(x,z)=K_{\mu\nu}\textrm{e}^{ipx}$ given previously.
In order to localize the 4D graviton, $\Psi_{0}(z)$
should obey the normalization condition
$\int dz |\Psi_{0}|^{2} < \infty$.
Generally, the character of the localization of massive KK modes
depends on the specific expression of the potential $V_{QM}$ of the
Schr\"{o}dinger equation. In the
following sections, we will investigate the mass spectrum and
the corrections of the Newtonian potential on bent thick branes.

\section{Localization and mass spectrum of gravity on dS thick branes}
\label{SecdSbrane}

The general formulation of the metric for a dS thick
brane embedded in AdS$_{5}$-like space-time is
\begin{eqnarray}\label{ds_line}
ds^{2}=\textrm{e}^{2A(z)}\left[-dt^{2}+\textrm{e}^{2Ht}(dx_{1}^{2}+dx_{2}^{2}+dx_{3}^{2})+dz^{2}\right].
\end{eqnarray}
Here $H$ is the dS parameter and the 4D cosmological constant is $\Lambda_{4}=3H^{2}$.
%
%The field equations generated from the action (\ref{action}) with
%the 5D metric (\ref{ds_line}) can be reduced to the
%following coupled nonlinear differential equations:
%%\begin{subequations}\label{dS_EinsteinEq}%%%%%%%%%%%%%%%%%%%%%%%%%%%%%%%%%%%%%%%%%%%%%%%%%%%%%%%%%%%%%
%\begin{eqnarray}
%\phi'^{2}&=&\frac{3}{\kappa_{5}^{2}}(A'^{2}-A''-H^{2}),\label{dS_EinsteinEq_a} \\
%V(\phi)  &=&\frac{3}{2\kappa_{5}^{2}}
%             \textrm{e}^{-2A}(3H^{2}-3A'^{2}-A''), \label{dS_EinsteinEq_b}\\
%\frac{d V(\phi)}{d\phi}
%         &=& \textrm{e}^{-2A}(3A'\phi'+\phi'').\label{dS_EinsteinEq_c}
%\end{eqnarray}
%%\end{subequations}
%Note that these equations are not independent. So there are a series
%of solutions for these equations. The relationship between these
%equations was discussed in \cite{Csaki_NPB_2000}.
%Next, we consider some explicit solutions of these equations.

\subsection{Example 1} \label{SecdSbrane_1}

%%%%%%%%%%%%%%%%%%%%%%%%%%%%%%%%%%%%%%%%%%%%%%%%%�����Ǿ���������Wang An-Zhong PRD������

Firstly, we consider the following solution \cite{Wang_PRD}:
\begin{eqnarray}\label{wang_dsbrane}
  A(z) &=& -\delta \ln [\cosh \bar{z}], \\
  \phi(z)&=&\phi_{0}\arcsin[\tanh\bar{z}], \\
  V(\phi)&=& \frac{3(1+3\delta)H^{2}}{2\delta}\cos^{2(1-\delta)}\big( \frac{\phi}{\phi_{0}} \big),
\end{eqnarray}
where $\bar{z}={H z}/{\delta}$, $\delta$ is a constant which satisfies $0<\delta<1$, and
$\phi_{0}\equiv[3\delta(1-\delta)]^{1/2}$. In this solution, the
constant $\kappa_{5}^{2}$ is set as $\kappa_{5}^{2}=1$. Such a
solution describes a dS thick brane located at $z=0$, and the range
of the fifth dimension is $-\infty<z<+\infty$.

From Eq. (\ref{wang_dsbrane}),
the potential of the Schr\"{o}dinger-like equation can be expressed
as
\begin{eqnarray}\label{VQM_wang}
V_{QM}(z)=
   \frac{3H^{2}}{4\delta}\left[3\delta-(2+3\delta)\textrm{sech}^{2}\bar{z}
   \right].
\end{eqnarray}
Now we investigate the localization and mass spectrum of gravity KK modes.
From Eq. (\ref{VQM_wang}), we know that the potential has a minimum
(negative value) $-{3H^{2}}/{(2\delta)}$ at $z=0$ and a maximum
(positive value) ${9H^{2}}/{4}$ at $z\rightarrow\pm\infty$. If
we set $p=H/\delta$ and $q=1+3\delta/2$, the Schr\"{o}dinger equation (\ref{Schrodinger_eq}) with
the potential (\ref{VQM_wang}) becomes
\begin{eqnarray}\label{Schrodinger_eq_wang}
[-\partial_{z}^{2}-q(q-1)p^{2}\textrm{sech}^{2}(pz)]\Psi_{n}=E_{n}\Psi_{n},
\end{eqnarray}
where $E_{n}=m_{n}^{2}-{9\delta^{2}p^{2}}/{4}$.
The mass spectrum of the bound states is found to be
\cite{Liu0708,DiazLiuJCAP}
%\begin{eqnarray}\label{E_n}
%E_{n}=-p^{2}(q-1-n)^{2}
%\end{eqnarray}
%or
\begin{eqnarray}\label{m_n_wang}
m_{n}^{2}=n(3\delta-n) {H^{2}}/{\delta^{2}},
\end{eqnarray}
where $n$ is an integer and satisfies
 $0\leq n\leq {3}\delta/{2}$. Note that the energy for
$n=0$ or $m_{0}=0$ always belongs to the spectrum of the potential
(\ref{VQM_wang}) for $\delta > 0$. So it can be seen that, for
$0<\delta\leq\frac{2}{3}$, there is only one bound state (ground
state)
\begin{eqnarray}\label{groundstate_wang}
\Psi_{0}(z) \propto %=\sqrt{\frac{H\Gamma(\frac{1}{2}+\frac{3\delta}{2})}
            %{\delta\sqrt{\pi}~\Gamma(\frac{3\delta}{2})}}
            \textrm{sech}^{3\delta/2}(\bar{z})
\end{eqnarray}
with $m_{0}^{2}=0$, which is just the massless mode
and stands for the localized 4D graviton on the dS brane. It is
clear that there is no tachyonic graviton mode (with $m^2<0$). The
continuous spectrum starts at $m^{2}={9}H^{2}/{4}$ and the
corresponding KK modes asymptotically turn into plane waves, which
represent delocalized KK massive gravitons. When ${2}/{3}<
\delta <1 $, there are two bound states, one is the ground state
(\ref{groundstate_wang}), the other is the first-excited state
\begin{eqnarray}\label{firststate_wang}
\Psi_{1}(z)\propto
         \textrm{sech}^{3\delta/2}(\bar{z})~\sinh \bar{z}
\end{eqnarray}
with mass $m_{1}^{2}=(3\delta -1)H^{2}/\delta$. The continuous
spectrum also starts at $m^{2}={9}H^{2}/{4}$.

The continuous
massive states $\Psi_{m}(z)$ are expressed by a linear
combination of the associated Legendre functions, and the normalized form is read as
%\begin{eqnarray}\label{KKmodes}
%\Psi_{m}(z)= c_{1}\textrm{P}\left({3 \delta}/{2}, i M,
%                       \tanh \bar{z}\right)
%                       +
%             c_{2}\textrm{Q}\left({3\delta}/{2}, i M,
%                       \tanh \bar{z}\right),
%\end{eqnarray}
%where the parameter
%$M=\sqrt{\frac{m^{2}\delta^{2}}{H^{2}}-\frac{9\delta^{2}}{4}}$, and
%$c_1$, $c_2$ are $m$-dependent parameters. The wave functions
%$\Psi_{m}(z)$ of the continuum KK modes are normalized as plane
%waves, i.e., $\Psi_{m}(z\rightarrow\infty)$ should approach to plane
%waves:
%\begin{eqnarray}\label{PlaneWave}
%\Psi_{m}(z)= k_{1}\textrm{e}^{i \frac{M H z}{\delta}}
%            +k_{2}\textrm{e}^{-i \frac{M H z}{\delta}}.
%\end{eqnarray}
%Further, we chose the parameters $k_{1}=k_{2}=\frac{1}{2}$ to
%normalize the plane wave functions $\Psi_{m}(z)$, thus the
%parameters $c_{1}$ and $c_{2}$ can be fixed, and so (\ref{KKmodes})
%is reduced to
\begin{eqnarray}\label{KKmodes2}
 \Psi_{m}(z) \!\!\!&=&\!\!\! \frac{1}{2}\big[ \Gamma(1+iM)
                \textrm{P}\left({3\delta}/{2}, -iM, \tanh\bar{z}\right)
                \nonumber \\
             &&~+
             \Gamma(1-iM) \textrm{P}\left({3\delta}/{2}, iM, \tanh\bar{z}\right)
             \big].
\end{eqnarray}
These KK modes would have contribution to the effective Newtonian potential, which will be considered later.

\subsection{Example 2}
\label{SecdSbrane_2}
%%%%%%%%%%%%%%%%%%%%%%%%%%%%%%%%%%%%%%%%%%%%%%%%%%%%%%D. Bazeia PLB634��2006��

In this subsection, we turn to another example. The solution is
given by \cite{afonso_plb2006}
\begin{eqnarray}
A(z)   &=&-\frac{1}{2}\ln[sa^{2}(1+H^{2})\cosh^{2}(kz)], \label{afonso_dsbrane_A} \\
\phi(z)&=&{k z}/{b}, \label{afonso_dsbrane_B} \\
%V(\phi)&=& \frac{3}{4} a^2 (H^2+1) \bigg\{3 \big[H^2 (s-1)-1\big]
%             \cosh ^2(b \phi )+4(H^2+1)\bigg\},\\
V(\phi)&=& \frac{9}{4} a^2 (H^2+1) \big[H^2 (s-1)-1\big]
             \cosh ^2(b \phi )\nonumber\\
        &&       +3 a^2 (H^2+1)^2, \label{afonso_dsbrane_C}
\end{eqnarray}
where $k^{2}=\frac{1+H^{2}}{s}$,
$b=\pm\sqrt{\frac{2(1+H^{2})}{3[1+(1-s)H^{2}]}}$,
 $s$ and $a$ are real parameters. In this example, the constant
$\kappa_{5}^{2}$ is set as $\kappa_{5}^{2}=2$.
%This solution
%describes a dS thick brane located at $z=0$, and the range of the
%fifth dimension is $-\infty<z<+\infty$.

The potential of the KK modes (\ref{VQM}) in the corresponding Schr\"{o}dinger equation turns out to
be
\begin{eqnarray}\label{VQM_afonso}
V_{QM}=({3k^{2}}/{4})[3-5\textrm{sech}^{2}(k z)],
\end{eqnarray}
which is similar to (\ref{VQM_wang}).
This potential supports two bound states. The first one is the ground
state with $m_{0}^{2}=0$:
\begin{eqnarray}\label{groundstate_Afonso}
\Psi_{0}(z)=\sqrt{{2k}/{\pi}}~
             \textrm{sech}^{\frac{3}{2}}(kz).
\end{eqnarray}
The second one is an excited state with $m_{1}^{2}=2k^{2}$:
\begin{eqnarray}\label{firststate_Afonso}
\Psi_{1}(z)\propto \textrm{sech}^{\frac{3}{2}}(kz)\sinh(kz).
\end{eqnarray}
The continuous spectrum starts at $m^{2}={9k^{2}}/{4}$.
The continuous KK modes are similar to those in the first example and we do not give them here.

Note that, in the first solution, we cannot get a flat brane solution by taking the limit $H\rightarrow0$. While, the flat brane scenario can be recovered from the second dS brane solution in the same limit.
For both dS brane solutions, there exists a mass gap and the 4D graviton can be localized on the dS branes.

\section{Localization and mass spectrum of gravity on AdS thick branes}
\label{SecAdSbrane}

For the case of AdS thick branes, the line-element has the following
form:
\begin{eqnarray}\label{Ads_line}
ds^{2}=\textrm{e}^{2A(z)}[\textrm{e}^{2H
x_{3}}(-dt^{2}+dx_{1}^{2}+dx_{2}^{2})+dx_{3}^{2}+dz^{2}].
\end{eqnarray}
Here the 4D cosmological constant is $\Lambda_{4}=-3H^{2}$. Next, we will use some
examples to investigate the localization and mass spectrum of
gravity on AdS branes.

\subsection{Example 1}
\label{SecAdSbrane_1}
%%%%%%%%%%%%%%%%%%%%%%%%%%%%%%%%%%%%%%%%%%%%%%%%%%%%%Wang An-zhong AdS brane

As the first example of AdS branes, we consider the following
solution \cite{Wang_PRD}:
\begin{eqnarray}\label{wang_Adsbrane}
A(z)&=&-\delta\ln \cos\bar{z}, \\
\phi(z)&=& \phi_{0} \textrm{arcsinh}\left[\tan\bar{z}\right], \\
V(\phi)&=&
-\frac{3(1+3\delta)H^{2}}{2\delta}\cosh^{2(1-\delta)}(\frac{\phi}{\phi_{0}}),
\end{eqnarray}
where $\phi_{0}\equiv\sqrt{3\delta(\delta-1)}$, and the constant
$\delta$ satisfies $\delta > 1$ or $\delta < 0$. In this example,
the constant $\kappa_{5}^{2}$ is set as $\kappa_{5}^{2}=1$. Such a
solution describes an AdS brane located at $z=0$, and the range of
the fifth dimension is $-z_{m}\leq z \leq +z_{m}$ with
$z_{m}=\left|\frac{\delta\pi}{2H}\right|$.

The potential of the KK modes reads as
\begin{eqnarray}\label{VQM_wang_Ads}
V_{QM}=\frac{3H^{2}}{4\delta}\left[(2+3\delta)\sec^{2}\bar{z}-3\delta\right].
\end{eqnarray}
It is clear that $V_{QM}(z=0)={3H^{2}}/{(2\delta)}$. At
the boundaries of the brane, the
potential $V_{QM}$ tends to positive infinite and negative infinite
for $\delta <-\frac{2}{3}$ or $\delta >1$ and
$-\frac{2}{3}<\delta<0$, respectively. So the mass gap always exists in the spectrum of gravitational
excitations of this system.

The case of $\delta>1$ is not interesting because the zero mode does not exist.
So we consider the case of $\delta<0$, for which the solutions of the KK modes are found to be
\begin{eqnarray}
\Psi_{n}(z) \!\!\! &=& \!\!\! c_n~
      _{2}\textrm{F}_{1}\left[-n, n-3\delta,
       ({1-3\delta})/{2}, ({1-\sin\bar{z}})/{2} \right]
      \nonumber \\
     && \quad
        \times \cos^{-\frac{3\delta}{2}}(\bar{z})\quad\quad
        (n=0,1,2,\cdots).  \label{PsiN_WangAdS}
\end{eqnarray}
All the KK modes are bound states and
can be localized on the thick AdS brane. The mass spectrum of the KK
modes is discrete:
\begin{eqnarray}\label{m_n_wang_AdS1}
m_{n}=\left|{H}/{\delta}\right|\sqrt{n(n+3|\delta|)}.
\end{eqnarray}
%Note that the ground state is just the zero mode
%\begin{eqnarray}\label{zeromode_wang_Ads}
%\Psi_{0}(z)\propto \textrm{e}^{\frac{3}{2}A}=
%     \cos(\frac{Hz}{\delta})^{-\frac{3\delta}{2}}.
%\end{eqnarray}

\subsection{Example 2}
\label{SecAdSbrane_2}

%%%%%%%%%%%%%%%%%%%%%%%%%%%%%%%%%%%%%%%%%%%%%%%D. Bazeia AdS brane

Another
solution of AdS branes  can be obtained by making replacement $H^{2}\rightarrow -H^{2}$ from (\ref{afonso_dsbrane_A})-(\ref{afonso_dsbrane_C}) \cite{afonso_plb2006}.  This solution describes an AdS thick
brane located at $z=0$, and the range of the fifth dimension is
$-\infty<z<+\infty$. The gravitation localizing potential is the same as (\ref{VQM_afonso}), so we do not need to repeat the analysis again.

%
%:
%\begin{eqnarray}
%A(z)   &=& -\frac{1}{2}\ln[sa^{2}(1-H^{2})\cosh^{2}(kz))], \label{afonso_Adsbrane_a} \\
%\phi(z)&=& \frac{k z}{b}, \label{afonso_Adsbrane_b}\\
%V(\phi)&=& \frac{9}{4} a^2(1-H^{2}) \big[H^2 (1-s)-1\big]
%             \cosh ^2(b \phi )\nonumber\\
%        &&   +3a^2(1-H^2)^2, \label{afonso_Adsbrane_c}
%\end{eqnarray}
%where $k^{2}=\frac{1-H^{2}}{s}$,
%$b=\pm\sqrt{\frac{2(1-H^{2})}{3[1-(1-s)H^{2}]}}$, $s$ and $a$ are
%real parameters, besides $s$ and $H^{2}$ should satisfy
%$s(1-H^{2})>0$. In this example, the constant $\kappa_{5}^{2}$ is
%set as $\kappa_{5}^{2}=2$. This solution describes an AdS thick
%brane located at $z=0$, and the range of the fifth dimension is
%$-\infty<z<+\infty$.
%
%By considering
%(\ref{afonso_Adsbrane_a}), the potential (\ref{VQM})
%reads
%\begin{eqnarray}\label{VQM_afonso_AdS}
%V_{QM}=\frac{3k^{2}}{4}[3-5\textrm{sech}^{2}(k z)].
%\end{eqnarray}
%Noting that the potential (\ref{VQM_afonso_AdS}) is the same as (\ref{VQM_afonso}), we do not need to consider it again.

\section{The effective Newtonian potentials}
\label{SecNewton}

In the brane world scenario, we need
to obtain the 4D effective action from the 5D
action (\ref{action}), i.e.,
\begin{eqnarray}
S_{5}  \supset M_{*}^{3} \int d^{5}x\sqrt{-g}R_{5}
           \supset  M_{Pl}^{2} \int d^{4}x\sqrt{-\hat{g}}\hat{R}_{4},
\end{eqnarray}
where $M_{*}=(2\kappa_5^2)^{-\frac{1}{3}}$ is the fundamental
5D Planck scale, and
$M_{Pl}=(M_{*}^{3}\int\textrm{e}^{3A(z)}dz)^{1/2}$ is the
4D Planck scale which determines the 4D
gravitational coupling $G_{N}\sim M_{Pl}^{-2}$. In other words, we
can get the 4D effective theory of gravity. Generally,
the localized zero mode will cause a 4D Newtonian
interaction potential, and we also require that the other KK modes
do not lead to unacceptably large corrections to the Newtonian
gravitational potential in 4D effective theory
\cite{rs,ADD,Csaki_NPB_2000,PRL845928,Brandhuber_JHEP}. In
this section, we will study the effective
Newtonian potentials on dS and AdS thick branes.

In the realistic brane scenario, the matter fields in the
4D theory on the brane would be smeared over the width of
the brane in the transverse space, which is too complex to deal
with. For simplicity, we just consider the gravitational potential
between two point-like sources of mass $M_{1}$ and $M_{2}$ located
at the location of the brane $z=0$
\cite{rs,ADD,Csaki_NPB_2000,Brandhuber_JHEP}. This assumption is
justified when the thickness of the brane is small compared with the
bulk curvature. Now, we consider two examples for dS and AdS branes.
For the four examples given above, the potentials $V_{QM}$ have only two types, i.e.,
the modified P\"{o}schl-Teller potential and infinite potential.
So, the discussion below for the effective Newtonian potentials
will focus on the first examples of dS and AdS branes.

\subsection{dS branes}

Firstly, we study the first example for the dS brane given above.
From Eq. (\ref{m_n_wang}), we know that, there are one and two bound states for
$0<\delta<{2}/{3}$ and ${2}/{3}<\delta<1$, respectively.
Thus we will discuss the two cases of $\delta$.

\textbf{Case I: $0<\delta<\frac{2}{3}$}

\begin{figure}[htb]
\begin{center}
\includegraphics[width=4cm]{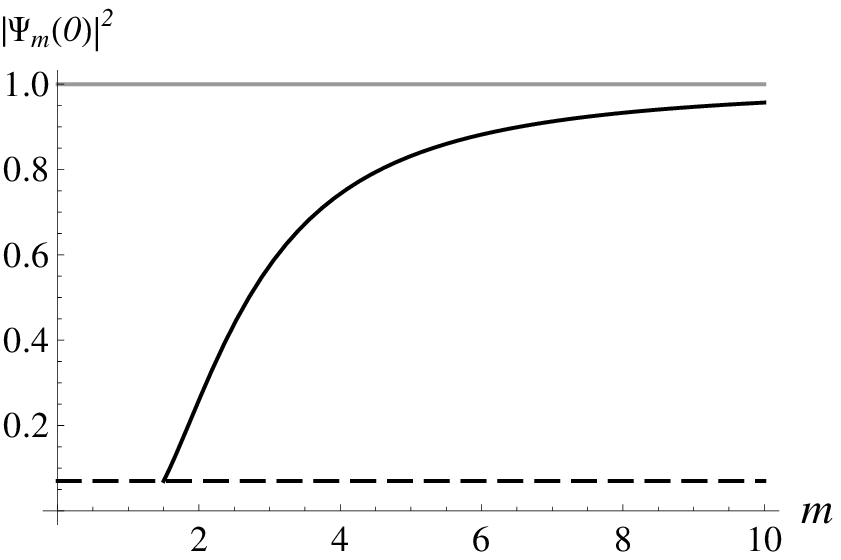}
\includegraphics[width=4cm]{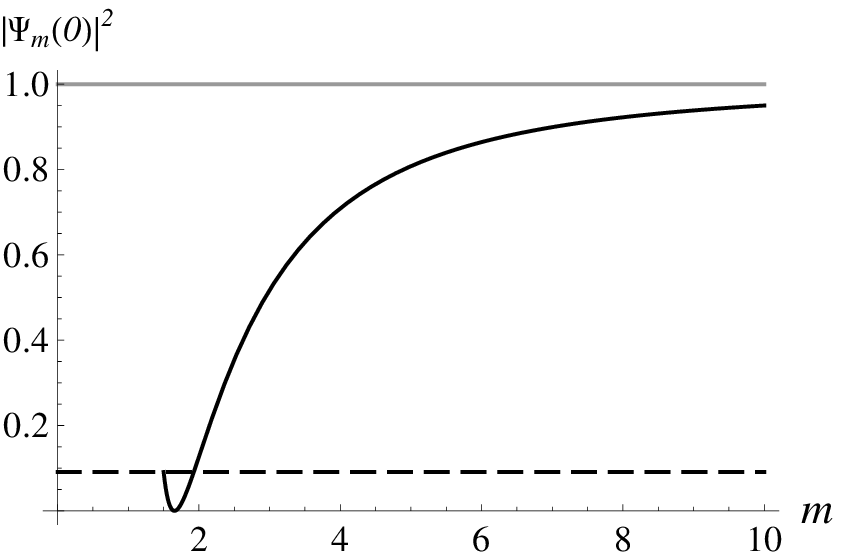}
\end{center}\vskip -3mm
\caption{The shape of $|\Psi_{m}(0)|^{2}$ (black line) as the function of $m$.
The black dashed line and gray line
represent $|\textrm{P}(\frac{3\delta}{2},0)|^{2}$ and the constant $1$, respectively.
The parameters are set as $H=1$, $\delta=0.5$ (left) and $\delta=0.9$ (right).} \label{fig_Psim1}
\end{figure}

In this case, the zero mode has been written as
(\ref{groundstate_wang}), and the continuum KK modes start at
$m_{c}={3H}/{2}$. The effective potential between two point-like
sources of mass $M_1$ and $M_2$ is from the contributions of the
zero mode and the continuum KK modes, and can be expressed as
\cite{Csaki_NPB_2000}
\begin{eqnarray}\label{EffectNewton}
U(r)=G_{N}\frac{M_{1}M_{2}}{r}+\frac{M_{1}M_{2}}{M_{*}^{3}}
        \int_{m_{c}}^{\infty}dm
        \frac{\textrm{e}^{-mr}}{r}|\Psi_{m}(0)|^{2},
\end{eqnarray}
where $G_{N}=1/(16{\pi}M_{Pl}^{2})$ is the Newton's gravitational constant.
The standard Newtonian
potential is from the contribution of the zero mode, and the
second term is the correction of the Newtonian potential, which is
from the contribution of the continuum KK modes (\ref{KKmodes2}).
For $m=m_{c}={3 H}/{2}$, we have
$\Psi_{m_{c}}(0)=\textrm{P}({3\delta}/{2},0)$. In Fig.
\ref{fig_Psim1}, the curve of $|\Psi_{m}(0)|^2$ as a function of $m$
is plotted. It is shown that
$|\Psi_{m_{c}}(0)|^{2}\leq|\Psi_{m}(0)|^2$ and $|\Psi_{m}(0)|^2$
approaches to the constant $1$ with the increase of the mass $m$,
this is because the wave functions $\Psi_{m}(z)$ approach to plane
waves as the eigenvalue $m$ becomes large.

%%%%%%%%%%%%%%%%%%%%%%%%%%%%%%%%%%%%%%%%%%%%%%%%%%%%%%%%%%%%%%%%%%%%%%%%%%
%%%%%%%%%%%%%%%%%%%%%%%%%%%%%%%%%%%%%%%%%%%%%%%%%%%%%%%%%%%

In order to get the corrected Newtonian potential, we write the
integrand as
%\begin{eqnarray}\label{Im}
$I(m)=({\textrm{e}^{-mr}}/{r})|\Psi_{m}(0)|^{2}$.
%\end{eqnarray}
Rather than dealing with this complex integrand, we will consider other
simple integrands, from which the integral can be achieved.

Firstly, we use the minimum $\Psi_{m_{c}}(0)$ to replace the
function $\Psi_{m}(0)$, thus the integrand can be simply given by
%\begin{eqnarray}
$I_{1}(m)=({\textrm{e}^{-mr}}/{r})\left|\textrm{P}({3\delta}/{2},0)\right|^{2}$.
%\end{eqnarray}
Because the range of the parameter $\delta$ is
$0<\delta<{2}/{3}$, the range of
$\textrm{P}({3\delta}/{2},0)$ is
$0<\Psi_{m_{c}}(0)=\textrm{P}({3\delta}/{2},0)<1$. In this case,
the correction of the Newtonian potential would be
\begin{eqnarray}\label{EffectNewtonCaseI}
\Delta U_{1}=\left|\textrm{P}(\frac{3\delta}{2},0)\right|^{2}
              \frac{\textrm{e}^{-\frac{3Hr}{2}}}{M_{*}^{3}}
               \frac{M_{1}M_{2}}{r^{2}}
             = \eta_{1}\frac{\textrm{e}^{-\frac{3Hr}{2}}}{M_{*}^{3}}
               \frac{M_{1}M_{2}}{r^{2}},
\end{eqnarray}
where $\eta_{1}=|\textrm{P}(\frac{3\delta}{2},0)|^{2}$.

Secondly, we use the constant $1$ to replace $|\Psi_{m}(0)|^{2}$,
which refers to the limit $|\Psi_{m\rightarrow\infty}(0)|^{2}=1$,
and the integrand can be simply given by $
I_{2}(m)={\textrm{e}^{-mr}}/{r}$.
Then the correction of the Newtonian potential would be
\begin{eqnarray}\label{EffectNewtonCaseI}
\Delta U_{2}&=& \frac{\textrm{e}^{-\frac{3Hr}{2}}}{M_{*}^{3}}
               \frac{M_{1}M_{2}}{r^{2}}.
\end{eqnarray}

From Fig. \ref{fig_Psim1}, we can see that
$\eta_{1}\leq|\Psi_{m}(0)|^{2}<1$, so the integrands satisfy the
relation $I_{1}\leq I_{m}<I_{2}$. Thus the correction of the
Newtonian potential $\Delta U(r)$ must satisfy
 $\Delta U_{1}(r)\leq\Delta U(r)<\Delta U_{2}(r)$ and the effective
Newtonian potential can be written as
\begin{eqnarray}\label{EffectNewP1}
U(r)=G_{N}\frac{M_{1}M_{2}}{r}+
     \eta_1\frac{\textrm{e}^{-\frac{3Hr}{2}}}{M_{*}^{3}}\frac{M_{1}M_{2}}{r^{2}}
\end{eqnarray}
with the constant $\eta$ satisfies
$|\textrm{P}({3\delta}/{2},0)|^{2}<\eta_1<1$.

\textbf{Case II: $\frac{2}{3}\leq\delta<1$}

In this case, the ground state $\Psi_{0}(z)$ (the zero mode) and the
first excited state $\Psi_{1}(z)$ are given by
(\ref{groundstate_wang}) and (\ref{firststate_wang}), respectively.
The continuum KK modes also start at $m_{c}={3H}/{2}$. The
effective potential between two point-like sources of mass $M_{1}$
and $M_{2}$ can be expressed as follows \cite{Csaki_NPB_2000}:
\begin{eqnarray}\label{EffectNewton2}
U(r)&=&G_{N}\frac{M_{1}M_{2}}{r}
     +  \frac{M_{1}M_{2}}{M_{*}^{3}}
     \frac{\textrm{e}^{-m_{1}r}}{r}\left|\Psi_{1}(0)\right|^{2} \nonumber\\
     && ~~~  +
        \frac{M_{1}M_{2}}{M_{*}^{3}}
        \int_{m_{c}}^{\infty}dm
        \frac{\textrm{e}^{-mr}}{r}|\Psi_{m}(0)|^{2},
\end{eqnarray}
where $m_{1}=\sqrt{(3\delta-1)H^{2}/\delta}$. Because $\Psi_{1}(z)$
is an odd function of $z$,
$\Psi_{1}(0)=0$ and Eq. (\ref{EffectNewton2}) can be reduced to Eq.
(\ref{EffectNewton}). The continuous massive states $\Psi_{m}(z)$
are also given by (\ref{KKmodes2}).
The curve of $|\Psi_{m}(0)|^{2}$ as a function of $m$ is plotted in Fig.
\ref{fig_Psim1}. We can see that $0\leq |\Psi_{m}(0)|^{2}<1$.
Following the same procedure for the first case, we can obtain the
effective Newtonian potential:
\begin{eqnarray}\label{EffectNewP2}
U(r)=G_{N}\frac{M_{1}M_{2}}{r}+
     \eta_2\frac{\textrm{e}^{-\frac{3Hr}{2}}}{M_{*}^{3}}\frac{M_{1}M_{2}}{r^{2}},
\end{eqnarray}
where $\eta_2$ is a constant satisfying $0<\eta_2<1$.

From above discussion, it can be seen that, for the large distance
$r$ of two point-like sources, the correction of gravitational
potential $\textrm{e}^{-\frac{3Hr}{2}}/\frac{1}{r^{2}}$ is a high
order term compared with the term of the Newtonian potential
$\frac{1}{r}$ and hence can be neglected. However, when $r$ is small, the
correction $\textrm{e}^{-\frac{3Hr}{2}}/r^{2}\sim\frac{1}{r^{2}}$
will be large and will be the main term of the effective Newtonian
potential. The conclusions for the second examples for dS and AdS branes are similar to this
one.

\subsection{AdS branes}

In this subsection, we investigate the first example for
AdS branes with $\delta<0$, for which all the KK modes including the zero mode are bound states. The localized zero mode will cause a 4D
Newtonian potential, and all other KK modes will produce the
correction to the potential. The effective potential of two
point-like sources is read as \cite{Csaki_NPB_2000}
\begin{eqnarray}\label{EffectNewtonCaseII}
U(r)=G_{N}\frac{M_{1}M_{2}}{r}+\sum_{n=1}^{\infty}
          \frac{M_{1}M_{2}}{M_{*}^{3}}\frac{\textrm{e}^{-m_{n}r}}{r}
           |\Psi_{n}(0)|^{2}.
\end{eqnarray}
Because all the odd-parity wave functions vanish at $z=0$, we just only
consider all even-parity wave functions. The even-parity wave functions are given by (\ref{PsiN_WangAdS})
but with $n=2i$, from which $\Psi_{n}(0)=\Psi_{2i}(0)$ read as
\begin{eqnarray}\label{KKmodezero}
\Psi_{2i}(0)=c_{2i}\frac{\sqrt{\pi}~\Gamma(\frac{1}{2}-\frac{3\delta}{2})}
                  {\Gamma(\frac{1}{2}-i)
                    \Gamma(\frac{1}{2}+i-\frac{3\delta}{2})}.
\end{eqnarray}

We introduce a new function $I(r)$ as follows
\begin{eqnarray}\label{Ir}
I(r)=\sum_{i=1}^{\infty}\frac{\textrm{e}^{-m_{2i}r}}{r}
           |\Psi_{2i}(0)|^{2},
\end{eqnarray}
where $m_{2i}=\left|\frac{H}{\delta}\right|\sqrt{2i(2i+3|\delta|)}$.
Then the correction of the Newtonian potential can be written as
%\begin{eqnarray}\label{deltaU_AdS}
$\Delta U(r)= ({M_{1}M_{2}}/{M_{*}^{3}}) I(r)$.
%\end{eqnarray}
Firstly, we need calculate the normalization constants
$c_{2i}$.
%\begin{eqnarray}
%c_{i}^{2}=\frac{1}{\int_{-z_{m}}^{+z_{m}}\cos^{-3\delta}(\frac{Hz}{\delta})
%              ~~\left[_{2}\textrm{F}_{1}\left(-2i,~
%              2i-3\delta,~
%              \frac{1-3\delta}{2},~
%              \frac{1-\sin^{2}(\frac{Hz}{\delta})}{2}\right)\right]^{2}dz}.
%\end{eqnarray}
The integrals are very complex, and we cannot give them.
Although the analytic expressions for $c_{2i}$ can not be obtained,
we can fix the parameters $\delta=-1$ and
$H=1$ and give the numerical results:
$ |\Psi_{2i}(0)|^{2}=$ 0.65625, 0.644531, 0.640869, 0.639267, 0.638426,
      0.63793,  0.637614, 0.637399,   0.637248, 0.637136, 0.637052, 0.636987,
      0.636935, 0.636894, 0.63686, $\cdots$.
We find that $|\Psi_{2i}(0)|^{2}$ are close to a constant for
different values of the parameter $i$. In the following calculation,
we use the constant $|\Psi_{2}(0)|^{2}$ to replace with all of
$|\Psi_{2i}(0)|^{2}$. Thus we obtain the following approximate expression
\begin{eqnarray}
I_{\texttt{app}}(r) &\approx&  |\Psi_{2}(0)|^{2}
\sum_{i=1}^{\infty}\frac{\textrm{e}^{-\left|\frac{H}{\delta}\right|r\sqrt{2i(2i+3|\delta|)}}}{r}
  \nonumber \\
   &\approx&  |\Psi_{2}(0)|^{2}
\sum_{i=1}^{\infty}\frac{\textrm{e}^{-\left|\frac{H}{\delta}\right|2ri}}{r}
   =  \frac{|\Psi_{2}(0)|^{2}}{(\textrm{e}^{2\left|\frac{H}{\delta}\right|r}-1)r}.~~~~~~~~
  \label{Ir_appro}
\end{eqnarray}

%%%%%%%%%%%%%%%%%%%%%%%%%%%%%%%%%%%%%%%%%%%%%%%%%%%%%%%%%%%%%%%%%%%%%%%%%%
\begin{figure}[htb]
\begin{center}
\includegraphics[width=6cm]{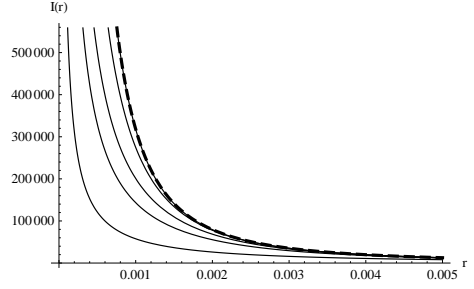}
\end{center}\vskip -5mm
\caption{The dashed line represents the shape of the approximate
expression $I_{\texttt{app}}(r)$ as the function of $r$. The solid lines represent
the shapes of $I_{N}(r)$ for different $N$. From left to right the
parameter $N$ is set to be $100, 300, 500, 1000, 5000$ respectively.
The other parameters are set as $H=1$ and $\delta=-1$.} \label{fig_Ir}
\end{figure}
%%%%%%%%%%%%%%%%%%%%%%%%%%%%%%%%%%%%%%%%%%%%%%%%%%%%%%%%%%%

Next, we need compare the approximate expression (\ref{Ir_appro})
with the exact result (\ref{Ir}). We introduce a new function
%\begin{eqnarray}
$I_{N}(r)=\sum_{i=1}^{N} ({\textrm{e}^{-m_{2i}r}}/{r})
           |\Psi_{2i}(0)|^{2}$.
%\end{eqnarray}
In Fig. \ref{fig_Ir}, we have plotted the shape of the approximate
expression $I_{\texttt{app}}(r)$ and the shapes of $I_{N}(r)$ for different values
of $N$. It can be seen that with the increase of $N$, $I_{N}(r)$
would tend to the approximate expression $I_{\texttt{app}}(r)$. Thus our expression
(\ref{Ir_appro}) is a good approximation. Then the effective
Newtonian potential can be expressed as:
\begin{eqnarray}\label{NewtonAdS}
U(r)=G_{N}\frac{M_{1}M_{2}}{r}
           +\frac{M_{1}M_{2}}{M_{*}^{3}}
           \frac{|\Psi_{2}(0)|^{2}}{(\textrm{e}^{2\left|\frac{H}{\delta}\right|r}-1)r}.
\end{eqnarray}
So the effective potential $U(r)$
can be reduced to the Newtonian potential for large distance $r$.
However, when the distance $r$ is small, then the correction term
for the gravitational potential $\Delta
U(r)\sim{\left(\textrm{e}^{2\left|\frac{H}{\delta}\right|r}-1\right)^{-1}/r}\sim
{1}/{r^{2}}$ will become the leading term.

From these two examples, we find that, when $r$ is large enough, the
correction of the gravitational potential can be ignored comparing
with the Newtonian potential; however, when $r$ is small, the
correction term will be the main term of the potential. The
corrections for the Newtonian potential for two different types of
localization potentials are similar at short distance: $\Delta
U(r)\sim {1}/{r^{2}}$.

\section{Conclusions and discussions}
\label{SecConclusion}

In this letter, the localization and mass spectrum of gravity, as
well as the effective Newtonian potentials for dS and AdS thick
branes have been investigated by two examples, respectively. We know
that the dS/AdS space-time would become the flat one if the dS/AdS
parameter $H=0$. Here, for the first examples of dS and AdS branes,
the flat brane solutions can not be got in the limit of
$H\rightarrow0$. While, the dS and AdS branes would turn to flat
ones for the second examples in the same limit. For all those bent
brane solutions, there exist mass gaps and the 4D graviton can be
localized on the  branes (for the first example of AdS brane, we
need $\delta<0$). The massive KK modes were obtained and the
effective gravitational potentials on the dS and AdS branes were
calculated. It was found that, for two different types of the
potentials of the gravity KK modes (a modified P\"{o}schl-Teller
potential and an infinite deep potential), the corrections of the
Newtonian potentials at short distance for two point-like sources
are the same: $\Delta U(r)\sim \frac{1}{r^{2}}$, which is different
from the correction in the RS brane model. The corrections can be
neglected when the distance between the two sources is large.

%\section{Section title}
%Insert here the text.
%See fig.~\ref{fig.1}, table~\ref{tab.1} and eq.~(\ref{eq.1}).
%See also~\cite{b.a,b.b}.
%\begin{equation}
%\label{eq.1}
%0\neq1
%\end{equation}
%
%
%\begin{figure}
%\onefigure{epl-template.eps}
%\caption{Figure caption.}
%\label{fig.1}
%\end{figure}
%
%
%\begin{table}
%\caption{Table caption.}
%\label{tab.1}
%\begin{center}
%\begin{tabular}{lcr}
%first  & table & row\\
%second & table & row
%\end{tabular}
%\end{center}
%\end{table}
%

\acknowledgments
This work was supported by the Huo Ying-Dong Education Foundation of
Chinese Ministry of Education (No. 121106), the National Natural
Science Foundation of China (No. 11075065),  and the Fundamental Research Funds for the Central Universities (No. lzujbky-2012-k30).


\begin{thebibliography}{0}

%\bibitem{b.a}
%  \Name{Author F., Author S. \and Author T.}
%  \REVIEW{Some Rev. A}{69}{1969}{9691}.
%
%\bibitem{b.b}
%  \Name{Author F. \and Author S.}
%  \Book{Some Book of Interest}
%  \Editor{A. Editor}
%  \Vol{9}
%  \Publ{Publishing house, City}
%  \Year{1939}
%  \Page{666}.
%
%\bibitem{b.c}
%  \Editor{Editor A.}
%  \Book{Some Book of Interest}
%  \Vol{9}
%  \Publ{Publishing house, City}
%  \Year{1939}
%  \Section{A}.

\bibitem{RubakovPLB1983136}
 \Name{Rubakov V. A. \and Shaposhnikov M. E.}
    %{\em Do we live inside a domain wall?},
    \REVIEW{Phys. Lett. B} {125} {1983} {136};
 \Name{Squires E. J.}
   %{\em Dimensional reduction caused by a cosmological constant},
    \REVIEW{Phys. Lett. B} {167}{1986} {286};
 \Name{Randjbar-Daemi S. \and Wetterich C. }
   %%%{\em Kaluza-Klein solutions with noncompact internal spaces},
    \REVIEW{ Phys. Lett. B} {166} {1986} {65};
 \Name{Antoniadis I.}
   %%%{\em A possible new dimension at a few Tev},
    \REVIEW{ Phys. Lett. B} {246} {1990} 377.


\bibitem{rs}
 \Name{Randall L. \and Sundrum R.}
   %%%{\em A Large Mass Hierarchy from a Small Extra Dimension},
    \REVIEW{ Phys. Rev. Lett.} {83} {1999} {3370};
    %arxiv:hep-ph/9905221;
 %Randall L and Sundrum R  1999
   %%%{\em An alternative to compactification},
    \REVIEW{ Phys. Rev. Lett.} {83} {1999} {4690}.
    %arXiv:hep-th/9906064.

\bibitem{Lykken}
 \Name{Lykken J. \and Randall  L.}
   %%%{\em The Shape of Gravity},
    \REVIEW{ JHEP }  {0006} {2000} {014}.
    %arXiv:hep-th/9908076.



\bibitem{ADD}
 \Name{Arkani-Hamed N., Dimopoulos S. \and Dvali  G.}
   %%%{\em The hierarchy problem and new dimensions at a millimeter},
    \REVIEW{ Phys. Lett. B}  {429} {1998} {263};
    %arXiv:hep-ph/9803315;
 \Name{Antoniadis I., Arkani-Hamed N., Dimopoulos S. \and Dvali G.}
    %%%{\em New dimensions at a millimeter to a Fermi and superstrings at a TeV},
    \REVIEW{ Phys. Lett.  B} {436} {1998} {257}.
    %arXiv:hep-ph/9804398.


\bibitem{CosmConst}
 \Name{Arkani-Hamed N., Dimopoulos S., Kaloper N. \and Sundtrum R.}
  %%%{\em A small cosmological constant from a large extra dimension},
    \REVIEW{ Phys. Lett. B} {480} {2000} {193};
    %arXiv: hep-th/0001197;
 \Name{Kehagias A.}
   %%%{\em A conical tear drop as a vacuum-energy drain for the
   %%%{   solution of the cosmological constant problem}},
    \REVIEW{ Phys. Lett. B} {600} {2004} {133}.
    %arXiv:hep-th/0406025.


\bibitem{De_Wolfe_PRD_2000}
 \Name{DeWolfe O., Freedman D. Z., Gubser S. S. \and Karch A.}
   %%%{\em Modeling the fifth dimension with scalars and gravity},
    \REVIEW{ Phys. Rev. D } {62} {2000} {046008}.
    %arXiv:hep-th/9909134.


\bibitem{Gremm_2000}
 \Name{Gremm M.}
   %%%{\em Four-dimensional gravity on a thick domain wall},
    \REVIEW{ Phys. Lett. B} {478} {2000} {434};
    %arXiv:hep-th/9912060;
 \Name{Gremm M.}
   %%%{\em Thick domain walls and singular spaces},
   \REVIEW{ Phys. Rev. D} { 62} {2000} {044017};
    %arXiv:hep-th/0002040;
 \Name{Giovannini M.}
   %%%{\em Gauge-invariant fluctuations of scalar branes},
    \REVIEW{ Phys. Rev. D} { 64} {2001} {064023};
    %arXiv:hep-th/0106041;
 %\Name{Kobayashi S., Koyama K. \and Soda J.}
%   %%%{\em Thick brane worlds and their stability},
%    \REVIEW{ Phys. Rev. D} { 65} {2002} {064014}.

\bibitem{Csaki_NPB_2000}
 \Name{Csaki C., Erlich J., Hollowood T. \and Shirman Y.}
   %%%{\em Universal Aspects of gravity localized on thick branes},
    \REVIEW{ Nucl. Phys. B} {581} {2000} {309}.
    %arXiv:hep-th/0001033.

\bibitem{PRL845928}
 \Name{Gregory R., Rubakov V. A. \and Sibiryakov S. M.}
  %%%{\em Opening up Extra Dimensions at Ultralarge Scales},
    \REVIEW{ Phys. Rev. Lett.} {84} {2000} {5928}.
  % arXiv:hep-th/0002072


\bibitem{Wang_PRD}
 \Name{Wang A.}
   %%%{\em Thick de Sitter 3-Branes, Dynamic Black Holes and Localization of Gravity},
     \REVIEW{ Phys. Rev. D} {66} {2002} {024024}.
     %arXiv:hep-th/0201051

\bibitem{varios}
 \Name{Guerrero R., Melfo A. \and Pantoja N.}
   %%%{\em Selfgravitating domain walls and the thin wall limit},
    \REVIEW{ Phys. Rev. D} {65} {2002} {125010};
 \Name{Castillo-Felisola O., Melfo A., Pantoja N. \and Ramirez A.}
   %%%{\em Localizing gravity on exotic thick three-branes},
    \REVIEW{ Phys. Rev. D} { 70} {2004} {104029};
 \Name{Herrera-Aguilar A., Malagon-Morejon D., Mora-Luna R. R. \and Nucamendi U.}
    %%%{\em Aspects of thick brane worlds: 4D gravity localization,
     %%%{smoothness, and mass gap}},
   \REVIEW{ Mod. Phys. Lett. A} {25} {2010} {2089}.
    %arXiv:0910.0363[hep-th].


\bibitem{ThickBrane}
 \Name{Dzhunushaliev V., Folomeev V., Singleton D. \and Aguilar-Rudametkin S.}
    %%%{\em 6D thick branes from interacting scalar fields},
     \REVIEW{ Phys. Rev. D} {77} {2008} {044006};
     %arXiv:hep-th/0703043;
 \Name{Bazeia D.,Brito F. A. \and Gomes A. R.}   2004
   %%%{\em Locally Localized Gravity and Geometric Transitions},
    \REVIEW{ JHEP} {0411}{2004} {070};
    %arXiv:hep-th/0411088;
 \Name{Bazeia D. \and Gomes A. R.}
   %%%{\em Bloch Brane},
    \REVIEW{ JHEP} {0405} {2004} {012};
    %arXiv:hep-th/0403141;
 \Name{Liu Y.-X., Zhong Y. \and Yang K.}
   %%%{\em Scalar-Kinetic Branes },
    \REVIEW{ Europhys. Lett.} {90} {2010} {51001}.
    % arXiv:0907.1952[hep-th].

\bibitem{Liu0708}
 \Name{Randjbar-Daemi S. \and Shaposhnikov M.}
    %%%{\em Fermion zero-modes on brane-worlds},
     \REVIEW{ Phys. Lett. B} { 492} {2000} {361};
     %arXiv:hep-th/0008079;
 \Name{Koley R. \and Kar S.}
   %%%{\em Scalar kinks and fermion localisation in warped spacetimes},
    \REVIEW{ Class. Quant. Grav.} {22} {2005} {753};
    %arXiv:hep-th/0407158;
 \Name{Liu Y.-X., Zhang X.-H., Zhang L.-D. \and Duan Y.-S.}
   %%%{\em Localization of Matters on Pure Geometrical Thick Branes},
    \REVIEW{ JHEP} {0802} {2008} {067};
    %arXiv:0708.0065[hep-th];
 \Name{Davies R., George D. P. \and Volkas R. R.}
   %%%{{\em Standard model on a domain wall brane}},
    \REVIEW{ Phys. Rev. D} {77} {2008} {124038}.
    %arXiv:0705.1584[hep-ph].
 \Name{Liu Y.-X., Zhang L.-D., Zhang L.-J. \and Duan Y.-S.}
   %%%{\em {Fermions on Thick Branes in the Background of Sine-Gordon Kinks}},
    \REVIEW{ Phys. Rev. D} {78}, {2008} {065025}.
    %arXiv:0804.4553[hep-th].


\bibitem{GherghettaPRL2000}
 \Name{Gherghetta T. \and Shaposhnikov M.}
   %%%{\em Localizing gravity on a string-like defect in six dimensions},
    \REVIEW{ Phys. Rev. Lett.} {85} {2000} {240};
    %arXiv:hep-th/0004014;
 \Name{Neupane I. P.}
   %%%{\em Consistency of higher derivative gravity in the brane background},
    \REVIEW{ JHEP} {0009} {2000} {040};
    %arXiv:hep-th/0008190;
 \Name{Liu Y.-X., Zhao L. \and Duan Y.-S.}
   %%%{\em Localization of Fermions on a String-like Defect},
    \REVIEW{ JHEP} {0704} {2007} {097};
    %arXiv:hep-th/0701010;
 \Name{Liu Y.-X., Zhao L., Zhang X.-H. \and Duan Y.-S.}
   %%%{\em Fermions in Self-dual Vortex Background on a String-like Defect},
    \REVIEW{ Nucl. Phys. B} {785} {2007} {234};
    %arXiv:0704.2812[hep-th];
 \Name{Almeida C. A. S., Casana R., Ferreira M. M. \and Gomes A. R.}
    %%%{\em Fermion localization and resonances on two-field thick branes},
    \REVIEW{ Phys. Rev. D} {79} {2009} {125022};
    %arXiv:0901.3543[hep-th];
 \Name{Liu Y.-X., Li H.-T., Zhao Z.-H., Li J.-X. \and Ren J.-R.}
   %%%{\em Fermion Resonances on Multi-field Thick Branes},
    \REVIEW{ JHEP }  {0910} {2009} {091}.
    % arXiv:0909.2312[hep-th]

\bibitem{Liu0907.0910}
 \Name{Koley R., Mitra J. \and SenGupta S.}
   %%%{\em Fermion localization in a generalized Randall-Sundrum model},
    \REVIEW{ Phys. Rev. D} { 79} {2009} {041902(R)};
 \Name{Liu Y.-X., Yang J., Zhao Z.-H., Fu C.-E. \and Duan Y.-S.}
   %%%{\em Fermion Localization and Resonances on
        %%%{A de Sitter Thick Brane}},
     \REVIEW{ Phys. Rev. D} { 80} {2009} {065019};
    %  arXiv:0904.1785[hep-th];
 \Name{Liu Y.-X., Fu C.-E., Zhao L. \and Duan Y.-S.}
    %%%{\em Localization and Mass Spectra of Fermions on
        %%%{Symmetric and Asymmetric Thick Branes}},
     \REVIEW{ Phys.  Rev. D} {80} {2009} {065020}.
    % arXiv:0907.0910[hep-th].


\bibitem{ThickBraneWeyl}
 \Name{Arias O., Cardenas R. \and Quiros I.}
   %%%{\em Thick Brane Worlds Arising From Pure Geometry},
    \REVIEW{ Nucl. Phys. B} {643} {2002} {187};
    %arXiv:hep-th/0202130.
 \Name{Barbosa-Cendejas N. \and Herrera-Aguilar A.}
   %%%{\em 4D gravity localized in non $Z_2$--symmetric thick branes}
    \REVIEW{ JHEP} {0510} {2005} {101};
    %arXiv:hep-th/0511050.
 \Name{Barbosa-Cendejas N. \and Herrera-Aguilar A.}
   %%%{\em Localization of 4D gravity on pure geometrical thick branes},
    \REVIEW{ Phys. Rev. D} {73} {2006} {084022}.
    %arXiv:hep-th/0603184.
 \Name{Liu Y.-X., Zhang L.-D., Wei S.-W. \and Duan Y.-S.}
   %%%{{\em Localization and Mass Spectrum of Matters on Weyl Thick Branes}},
    \REVIEW{ JHEP} {0808} {2008} {041}.
    %arXiv:0803.0098[hep-th].

\bibitem{Cvetic}
 \Name{Cvetic M., Griffies S. \and Rey S.-J.}
      %%%{\em Static Domain Walls in N=1 Supergravity}
        \REVIEW{ Nucl. Phys. B} {381} {1992} {301};
         %arXiv:hep-th/9201007;
 \Name{Cvetic M. \and Soleng H. H.}
       %%%{\em Supergravity Domain Walls},
         \REVIEW{ Phys. Rept.} {282} {1997} {159};
         %arXiv:hep-th/9604090;
 \Name{Cvetic M. \and Robnik M.}
      %%%{\em Gravity Trapping on a Finite Thickness Domain Wall: An Analytic Study},
         \REVIEW{ Phys. Rev. D} {77} {2008} {124003}.
         %arXiv:0801.0801[hep-th].


\bibitem{PRD0709.3552}
 \Name{Barbosa-Cendejas N., Herrera-Aguilar A.,
  Reyes Santos M. A. \and Schubert C.}
   %%%{{\em Mass gap for gravity localized on Weyl thick branes}},
    \REVIEW{ Phys. Rev. D} {77} {2008} {126013}.
    %arXiv:0709.3552[hep-th].


 \bibitem{DiazLiuJCAP}
 \Name{Diaz J. I., Negro J., Nieto L. M., Rosas-Ortiz O.}
 %%%{\em The supersymmetric modified P\"{o}schl-Teller and delta well potentials },
       \REVIEW{ J. Phys. A: Math. Gen.} {32} {1999} {8447};
 \Name{Liu Y.-X., Zhao Z.-H., Wei S.-W. \and Duan Y.-S.}
   %%%{\em Bulk Matters on Symmetric and Asymmetric de Sitter Thick Branes},
   \REVIEW{ JCAP } {0902} {2009} {003}.
    %arXiv:0901.0782[hep-th].


\bibitem{RandallJHEP2001}
 \Name{Kobayashi S., Koyama K. \and Soda J.}
   %%%{\em Thick brane worlds and their stability},
    \REVIEW{ Phys. Rev. D} { 65} {2002} {064014};
 \Name{Karch A. \and Randall L.}
   %%%{\em Locally Localized Gravity},
    \REVIEW{ JHEP} {0105} {2001} {008}.




\bibitem{afonso_plb2006}
 \Name{Afonso V. I., Bazeia D. \and Losano L.}
    %%%{\em First-order formalism for bent brane},
     \REVIEW{ Phys. Lett. B} {634} {2006} {526}.
     %arXiv:hep-th/0601069

\bibitem{liu_0911.0269}
 \Name{Liu Y.-X., Yang K. \and Zhong Y.}
   %%%{\em de Sitter Thick Brane Solution in Weyl Geometry},
    \REVIEW{ JHEP} {1010} {2010} {069}.
    % arXiv:0911.0269[hep-th]


%\bibitem{1004.0150}
% \Name{Balcerzak A. \and D\c{a}browski M. P.}
%      %%%{\em Brane f(R) gravity cosmologies},
%       \REVIEW{ Phys. Rev. D} {81} {2010} {123527}.
%     %arXiv:1004.0150[hep-th]

\bibitem{Liu_2010}
 \Name{Li H.-T., Liu Y.-X., Zhao Z.-H. \and Guo H.}
    %%%{\em Fermion Resonances on a Finite Thickness Brane},
      \REVIEW{ Phys. Rev. D} {83} {2011} {045006};
      % arXiv:1006.4240[hep-th];
 \Name{Zhao Z.-H., Liu Y.-X., Li H.-T. \and Wang Y.-Q.}
    %%%{\em Temperature Effect on the Fermion Localization and Resonances on Thick ranes},
        \REVIEW{ Phys. Rev. D} {82} {2010} 084030;
       % arXiv:1004.2181[hep-th];
 \Name{Liu Y.-X., Fu C.-E., Guo H., Wei S.-W. \and Zhao Z.-H.}
     %%%{\em Bulk Matters on a GRS-Inspired Braneworld},
      \REVIEW{ JACP } {1012} {2010} {031};
       % arXiv:1002.2130[hep-th];
 \Name{Neupane I. P.}
 %%%{\em Warped compactification to de Sitter space},
  \REVIEW{ Nucl. Phys. B} {847} {2011} {549}.


%%%%%%%%%%%%%%%%%%%%%%%%%%%%%%%%%%%%%%%%%%%%%%%%%%%%%%%%%%% review
%\Name{Author F. \and Author S.}
%  \Book{Some Book of Interest}
%  \Editor{A. Editor}
%  \Vol{9}
%  \Publ{Publishing house, City}
%  \Year{1939}
%  \Page{666}.


\bibitem{reviews}
 \Name{Mannheim P. D.}
        \Book{Brane-localized Gravity},
        \Publ{World Scientific Publishing Company, Singapore}
        \Year{2005};
 \Name{SenGupta S.}
    {\em Aspects of warped braneworld models},
      {arXiv:0812.1092[hep-th]};
 \Name{Dzhunushaliev V., Folomeev V. \and Minamitsuji M.}
   %%%{\em Thick brane solutions},
    \REVIEW{ Rept. Prog. Phys.} {73} {2010} {066901};
    %arXiv:0904.1775[gr-qc];
 \Name{Shifman M.}  2010
   %%%{\em Large extra dimensions: Becoming acquainted with an
    %%%{ alternative paradigm}},
     \REVIEW{ Int. J. Mod. Phys. A} {25} {2010} {199};
    % arXiv:0907.3074[hep-ph];
 \Name{Rizzo T. G.}
    %%%{\em Introduction to extra dimensions},
     \REVIEW{ AIP Conf. Proc.} {1256} {2010} {27};
      % arXiv:1003.1698[hep-ph];
 \Name{Maartens R. \and Koyama K.}
       %%%{\em Brane-World Gravity},
        \REVIEW{ Living Rev. Rel.} {13} {2010} {5}.
        %arXiv:1004.3962 [hep-th].


%%%%%%%%%%%%%%%%%%%%%%%%%%%%%%%%%%%%%%%%%%%%%%%%%%%%%%%%


\bibitem{Brandhuber_JHEP}
 \Name{Brandhuber A. \and Sfetsos K.}
    %%%{\em  Non-standard compactifications with mass gaps and Newton's law},
     \REVIEW{ JHEP}  {9910} {1999} {013}.
     %arXiv:hep-th/9908116






\end{thebibliography}
\end{document}